\RequirePackage{fix-cm}
\documentclass[smallextended]{svjour3}       
\smartqed  
\usepackage{graphicx}
\usepackage{amsmath}
\usepackage{amssymb}
\usepackage{latexsym}
\usepackage{mathtools}
\usepackage{mathptmx}
\usepackage{nicefrac}
\usepackage{bm}
\usepackage[normalem]{ulem}
\usepackage{tabularx}

\usepackage[british,english]{babel}

\usepackage{array}
\usepackage{booktabs}

\usepackage{url}
\usepackage{cite}
\usepackage{xcolor}
\definecolor{newcolor}{rgb}{.8,.349,.1}

\usepackage{tikz}
\usepackage{pgfplots}
\pgfplotsset{compat=1.3}
\usepackage{filecontents}
\usetikzlibrary{matrix}
\usetikzlibrary{calc}
\usetikzlibrary{arrows}
\usetikzlibrary{patterns}
\usetikzlibrary{plotmarks}
\usetikzlibrary{intersections} 
\usetikzlibrary{decorations.pathmorphing}
\usetikzlibrary{decorations.pathreplacing}
\usetikzlibrary{decorations.markings,arrows} 
\usetikzlibrary{decorations.shapes}
\usetikzlibrary{fit}
\usetikzlibrary{backgrounds}
\usetikzlibrary{trees}

\usepackage{hyperref}

\usepackage[algoruled,vlined,linesnumbered]{algorithm2e}

\newcommand{\set}[1]{\left\{ #1 \right\}}
\newcommand{\seq}[1]{\left( #1 \right)}
\newcommand{\bmat}[1]{\mathbf{#1}}
\newcommand{\bvec}[1]{\bm{#1}}
\newcommand{\loss}[1]{\mathcal{L}\left(#1\right)}
\newcommand{\norm}[1]{\left\lVert#1\right\rVert}
\newcommand{\uniform}[2]{\mathrm{unif}\!\left\{#1, #2\right\}}

\newcommand{\BigO}[1]{\ensuremath{\operatorname{O}\!\left(#1\right)}}

\usepackage[colorinlistoftodos]{todonotes}
\usepackage{soul}

\journalname{Multimedia Tools and Applications}

\begin{document}

\title{Large-Scale User Modeling with Recurrent Neural Networks for Music Discovery on Multiple Time Scales}


\titlerunning{Large-Scale User Modeling with RNNs for Music Discovery}        

\author{Cedric De Boom		\and
		Rohan Agrawal		\and
		Samantha Hansen		\and
		Esh Kumar			\and
		Romain Yon			\and
		Ching-Wei Chen		\and
		Thomas Demeester		\and
		Bart Dhoedt		
}

\authorrunning{C.~De Boom, et al.} 

\institute{Cedric De Boom, Thomas Demeester, Bart Dhoedt \at
              Ghent University, imec, IDLab\\
       			Technologiepark 15\\
       			9052 Ghent, Belgium\\
       			\email{$\{$cedric.deboom, thomas.demeester, bart.dhoedt$\}$@ugent.be}
           \and
           Rohan Agrawal, Samantha Hansen, Esh Kumar, Romain Yon, Ching-Wei Chen \at
              Spotify, Inc.\\
       			45 W 18th St\\
       			10011 New York, NY, USA\\
       		\email{$\{$rohanag, slhansen, eshvk, romain, vidhya, cw$\}$@spotify.com}
}


\maketitle

\begin{abstract}
The amount of content on online music streaming platforms is immense, and most users only access a tiny fraction of this content.
Recommender systems are the application of choice to open up the collection to these users.
Collaborative filtering has the disadvantage that it relies on explicit ratings, which are often unavailable, and generally disregards the temporal nature of music consumption.
On the other hand, item co-occurrence algorithms, such as the recently introduced word2vec-based recommenders, are typically left without an effective user representation.
In this paper, we present a new approach to model users through recurrent neural networks by sequentially processing consumed items, represented by any type of embeddings and other context features.
This way we obtain semantically rich user representations, which capture a user's musical taste over time.
Our experimental analysis on large-scale user data shows that our model can be used to predict future songs a user will likely listen to, both in the short and long term.
\keywords{Recommender systems \and machine learning \and recurrent neural networks \and deep learning \and word2vec \and music information retrieval \and representation learning}
\end{abstract}

\section{Introduction}
\label{sec:introduction}
Online digital content providers, such as media streaming services and e-commerce websites, usually have immense catalogs of items.
To prevent users from having to search manually through the entire catalog, recommender systems help to filter out items users might like to watch, listen to, buy\dots~and are often based on characteristics of both users and items.
One type of recommendation algorithms is collaborative filtering, which is generally based on ratings for items given by users.
However, such explicit ratings are not always available.
For example, in what way does clicking on an item represent how much the user likes this item?
Implicit feedback models are therefore used here, but they require careful parameter tuning.
Next to this, systems that model users based on aggregate historical consumption will often ignore the notion of sequentiality.
In the case of music consumption, for example, it has been investigated that a user's listening behavior can be described by a trajectory in time along different artists and genres with periods of fixations and transitions \cite{Figueiredo:2016ut}. 


Recently, recommenders have also been built on top of item embeddings \cite{Barkan:2016wm, Liang:2016wj}.
Such embeddings, or vector representations, are generally learned using item co-occurrence measures inspired by recent advances in language modeling, e.g.~word2vec and related algorithms \cite{Mikolov:2013uz}.
The problem with this approach is that we are left without an adequate user representation, and the question remains how to derive such a representation based on the given item embeddings.

In this work we focus on creating user representations in the context of new music discovery on online streaming platforms.
We start from given latent embeddings of the items in the catalog and we represent users as a function of their item consumption.
For this, we propose that a user's listening behavior is a sequential process and is therefore governed by an underlying time-based model.
For example, think of a user listening to an artist's album for some time and then transitioning to the next album, or to a compilation playlist of the same musical genre.
To model the dynamics of a user's listening pattern we use recurrent neural networks, which are currently among the state-of-the-art in many data series processing tasks, such as natural language \cite{Sutskever:2014ty, Bahdanau:2015vz} and speech processing \cite{Graves:2014vz}.
We do not presuppose any of the item embedding properties, such that our model is generally applicable to any type of item representation.
In the next section we will explain the problem setting and highlight related work regarding music recommendation and deep learning.
In Section \ref{sec:methodology} we will describe our methodology, after which we perform an initial data analysis in Section \ref{sec:datagathering}. We conclude with the results of our experiments in Section \ref{sec:experiments}.
A complete table with the used symbols in this article is given in Appendix \ref{sec:appendix}.

\section{Motivation and Related Work}
\label{sec:motivation}
Ever since the launch of the Netflix Prize competition in 2006 \cite{Bennett:2007uv}, research in recommender systems, and a particular subset called collaborative filtering, has spiked.
The basis of modern collaborative filtering lies in latent-factor models, such as matrix factorization \cite{Koren:2009jg}.
Herein, a low-dimensional latent vector is learned for each item and user based on rating similarities, and in the most basic scheme the dot product between a user vector $\bvec{v}_u$ and item vector $\bvec{v}_i$ is learned to represent the rating $r_{ui}$ of item $i$ by user $u$: 
\begin{align}
r_{ui} = \bvec{v}_u^T\bvec{v}_i.
\end{align}
This setting is based on the entire user and item history, and does not take into account that a user's taste might shift over time.
Koren et al.~\cite{Koren:2009jg} mention that such temporal effects can easily be brought into the model by adding time-dependent biases for each item and user:
\begin{align}
r_{ui}(t) = \mu + b_u(t) + b_i(t) + \bvec{v}_u^T\bvec{v}_i(t).
\end{align}
Dror et al.~\cite{Dror:2011bb} extend on this work by introducing additional biases for albums, artists, genres and user sessions, but these biases now represent a global rather than a temporal effect of a user's preference towards an item.
This way we can for example model to what extent the average rating for a specific album is higher or lower compared to other albums.
Although the models of Koren et al.~and Dror et al.~are capable of representing a user's overall preference towards an item and how this preference shifts in time, it cannot immediately explain why a user would rate item $w$ higher or lower after having consumed items $x$, $y$ and $z$.
That is, a user's preference can depend on what he or she has consumed in the immediate past.
Basic collaborative filtering techniques do not explicitly model this sequential aspect of item consumption and the effect on what future items will be chosen.

Next to this, standard collaborative filtering and matrix factorization techniques are mostly fit for explicit feedback settings, i.e.~they are based on positive as well as negative item ratings provided by the users.
In more general cases where we deal with views, purchases, clicks\dots~we only have positive feedback signals, which are binary, non-discriminative, sparse, and are inherently noisy \cite{Hu:2008el}.
Standard and tested techniques to factorize a sparse matrix of user-item interactions are singular value decomposition (SVD) and non-negative matrix factorization (NMF) \cite{Anonymous:H3H5BbuI, Lee:2000ti}.
In the context of implicit feedback, however, missing values do not necessarily imply negative samples.
Pan et al.~\cite{Pan:2008kb} therefore formulate this as a so-called one class classification problem, in which observed interactions are attributed higher importance than non-observed ones through different weighting schemes, or through careful negative sampling.
Hu et al.~\cite{Hu:2008el} on the other hand construct an implicit matrix factorization algorithm, based on the singular value decomposition of the user-item matrix, which differs from the standard algorithm by attaching higher confidence on items with a large number of interactions during optimization.
Johnson \cite{Johnson:2014tf} uses the ideas from Hu et al.~and devises a probabilistic framework to model the probability of a user consuming an item using logistic functions.

The implicit-feedback models calculate global recommendations and do not exploit temporal information to decide which items the user might be interested in.
Recently, Figueiredo et al.~\cite{Figueiredo:2016ut} have shown that users on online music streaming services follow a trajectory through the catalog, thereby focusing their attention to a particular artist for a certain time span before continuing to the next artist.
For this they use a combination of different Markov models to describe the inter- (`switch') and intra-artist (`fixation') transitions.
In other work, Moore et al.~\cite{Moore:2013tj} learn user and song embeddings in a latent vector space and model playlists using a first-order Markov model, for which they allow the user and song vectors to drift over time in the latent space.

Compared to Markov models, which inherently obey the Markov property, recent work has shown that recurrent neural networks (RNNs) are able to learn long-term data dependencies, can process variable-length time series, have great representational power, and can be learned through gradient-based optimization. 
They can effectively model the non-linear temporal dynamics of text, speech and audio \cite{VanDenOord:2016uo, Sercu:2016ub, Karpathy:2015wu}, so they are ideal candidates for sequential item recommendation.
In a general RNN, at each time step $t$ a new input sample $x_t$ is taken to update the hidden state $h_t$:
\begin{align}
h_t = \mathcal{F} (Ux_t + Wh_{t-1}),
\end{align}
in which $\mathcal{F}(\cdot)$ is a non-linear function, e.g.~sigmoid $\sigma(\cdot)$, $\tanh$ or a rectifier (ReLU and variants) \cite{Maas:2013tn}.
To counter vanishing gradients during backpropagation and to be able to learn long-term dependencies, recurrent architectures such as long short-term memories (LSTMs) and gated recurrent units (GRUs) have been proposed, both with comparable performances \cite{Hochreiter:1997fq, Greff:2015wv, Chung:2014wf}.
These models use a gating mechanism, e.g.~an LSTM introduces input ($i_t$) and forget ($f_t$) gates that calculate how much of the input is taken in and to what extent the hidden state should be updated, and an output gate ($o_t$) that leaks bits of the internal cell state ($c_t$) to the output:
\begin{align}
i_t &= \sigma\left( U_ix_t + W_ih_{t-1} + w_i \odot c_{t-1} + b_i\right), \nonumber\\
f_t &= \sigma\left( U_fx_t + W_fh_{t-1} + w_f \odot c_{t-1} + b_f\right), \nonumber\\
c_t &= f_t \odot c_{t-1} + i_t \odot \tanh\left( U_cx_t + W_ch_{t-1} + b_c \right), \nonumber\\
o_t &= \sigma\left( U_ox_t + W_oh_{t-1} + w_o \odot c_{t-1} + b_o\right), \nonumber\\
h_t &= o_t \odot \tanh\left( c_t \right),
\end{align}
in which $\odot$ is the element-wise vector multiplication.
GRUs only have a reset ($r_t$) and update ($u_t$) gate, get rid of the cell state, and have less parameters overall:
\begin{align}
r_t &= \sigma\left( U_rx_t + W_rh_{t-1} + b_r\right), \nonumber\\
u_t &= \sigma\left( U_ux_t + W_uh_{t-1} + b_u\right), \nonumber\\
g_t &= \tanh\left( U_gx_t + r_t\odot W_gh_{t-1} + b_g \right), \nonumber\\
h_t &= (1 - u_t) \odot h_{t-1} + u_t \odot g_t.
\end{align}

Very recently there have been research efforts in using RNNs for item recommendation.
Hidasi et al.~\cite{Hidasi:2015uq} use RNNs to recommend items by predicting the next item interaction.
The authors use one-hot item encodings as input and produce scores for every item in the catalog, on which a ranking loss is defined.
The task can thus be compared to a classification problem.
For millions of items, this quickly leads to scalability issues, and the authors resort to popularity-based sampling schemes to resolve this.
Such models typically take a long time to converge, and special care needs to be taken not to introduce a popularity bias, since popular items will occur more frequently in the training data.
The work by Tan et al.~\cite{Tan:2016vy} is closely related to the previous approach, and they also state that making a prediction for each item in the catalog is slow and intractable for many items.
Instead, low-dimensional item embeddings can be predicted at the output in a regression task, a notion we will extend on in Section \ref{sec:methodology}.

A popular method to learn item embeddings is the word2vec suite by Mikolov et al.~\cite{Mikolov:2013uz} with both Continuous Bag-of-Words and Skip-Gram variants.
In this, a corpus of item lists is fed into the model, which learns distributed, low-dimensional vector embeddings for each item in the corpus.
Word2vec and variants have already been applied to item recommendation, e.g.~Barkan et al.~\cite{Barkan:2016wm} formulate a word2vec variant to learn item vectors in a set consumed by a user, Liang \cite{Liang:2016wj} devise a word2vec-based CoFactor model that unifies both matrix factorization and item embedding learning, and Ozsoy \cite{Ozsoy:2016tm} learns embeddings for places visited by users on Foursquare to recommend new sites to visit.
These works show that a word2vec-based recommender system can outperform traditional matrix factorization and collaborative filtering techniques on a variety of tasks.
In the work by Tan et al.~item embeddings are predicted and at the same time learned by the model itself, a practice that generally deteriorates the embedding quality: in the limit, the embeddings will all collapse to a degenerate, non-informative solution, since in this case the loss will be minimal.
Also, they minimize the cosine distance during training, which we found to decrease performance a lot.

In the coming sections, we will train RNNs to predict songs a user might listen to in the future as a tool to model users on online music streaming platforms.
For this, we will predict preexisting item embeddings---about which we will not make any assumptions in the model---and our task is therefore a regression rather than a classification problem.
This approach is closely related to the work by van den Oord et al.~\cite{VanDenOord:2013tp} in which collaborative latent vectors are predicted based on raw audio signals, and also related to the work by Hill et al.~\cite{Hill:2016wg} who learn to project dictionary definition representations onto existing word embeddings.
Regarding sequential item recommendation, the related works by Hidasi et al.~and Tan et al.~mentioned above both perform item recommendation within user sessions, i.e.~uninterrupted and coherent sequences of item interactions, which can last from ca.~10 minutes to over an hour.
Here, the prediction time scale is very short-term, and since consumed items within a user session are usually more similar than across user sessions, it is generally easier perform item recommendation on this short time scale.
In this work we will explore recommending songs for short-term as well as long-term consumption.
To recommend songs on the long term, we will need to be able to model a user's behavior across session boundaries.

\section{RNNs for Music Discovery}
\label{sec:methodology}
In this section we will explain the details of our approach to use RNNs as a means to model users on online music streaming platforms and to recommend new songs in a music discovery context.
Since we aim towards models that can be used efficiently in large-scale recommendation pipelines, we require the models to be trained within a reasonable time frame.
Furthermore, sampling from the model should be efficient.
Small models with little parameters are typically wanted to satisfy both requirements.

\begin{figure}[t!]
\includegraphics[width=.8\linewidth]{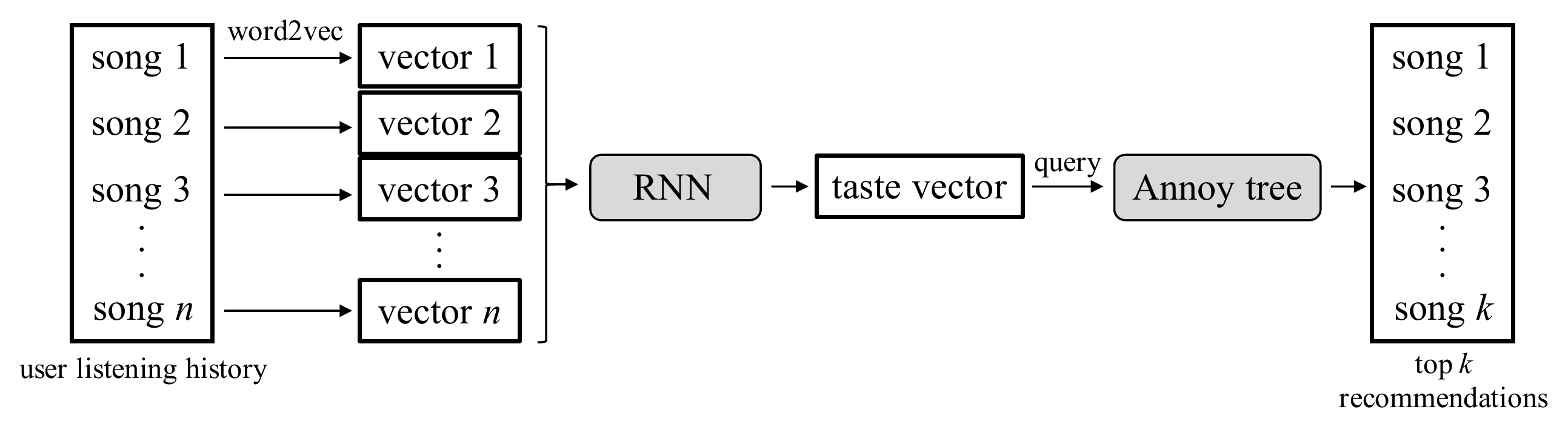}
\caption{The entire song recommendation pipeline for a specific user; we start with the user's listening history of $N$ songs, and we end the pipeline with $k$ song recommendations.}
\label{fig:overview}
\end{figure}

The entire recommendation pipeline for one specific user is given in Figure \ref{fig:overview}.
The basic building blocks of the pipeline are song vectors, which have been learned using the songs in the catalog.
The general idea is then to capture and predict a \textit{taste vector} for each user.
These taste vectors are the output of an RNN that sequentially aggregates song vectors from the user's listening history, and can therefore be regarded as a representation of the user's musical taste.
The taste vector can subsequently be used to generate song recommendations by querying a tree data structure for the nearby song vectors.

Since we construct real-valued taste vectors, the RNN solves a regression task rather than a classification task, as argued in Section \ref{sec:motivation}.
Directly predicting item embeddings is a regression problem that requires predicting a limited set of real-valued outputs, as opposed to a classifier with as many outputs as the number of items.
The computational footprint of these models is typically smaller than the classifiers.
They are usually learned faster, and are not per se biased towards popular items.
One of the main advantages is that any type of item embeddings and embedding combinations, along with other features, can be used to learn the regression model.

We break the recommendation pipeline into three separate stages which we will cover below.
First, we learn low-dimensional embeddings for each song in the streaming catalog using word2vec (\S\ref{subsec:embeddings}).
Then, we use an RNN to generate taste vectors for all users in the song embedding space (\S\ref{subsec:tastevectors}).
Finally, we use the taste vector to query songs in the nearby space to generate recommendations for all users (\S\ref{subsec:recommending}).

\subsection{Learning song embeddings}\label{subsec:embeddings}
In the first stage of the recommendation pipeline we learn latent vector representations for the top $N$ most popular songs in the catalog.
For this, we use Google's word2vec suite as explained in Section \ref{sec:motivation}; more specifically we use the Continuous Bag-of-Words (CBoW) algorithm with negative sampling.
As input to the word2vec algorithm we take user-created playlists of songs.
In this, each playlist is considered as an ordered `document' of songs.
By scanning all playlists in a windowed fashion, word2vec will learn a distributed vector representation $\bvec{s}$ with dimensionality $D$ for every song $s$.
Details regarding training data for word2vec will be highlighted in Section \ref{sec:datagathering}.

\subsection{Learning user taste vectors}\label{subsec:tastevectors}
In the second pipeline stage we use RNNs to produce user taste vectors based on song listening history.
The network takes a sequence of song vectors of dimensionality $D$ as input and produces a single taste vector with the same dimensionality $D$.
Let's denote the set of all users by $U$, the ordered sequence of song vectors user $u$ listened to by $\seq{\bvec{s}^u}$, and the predicted taste vector by $\bvec{t}^u$.
The RNN then produces:
\begin{equation}
\forall u \in U\colon \bvec{t}^u = \mathcal{R}\left( \seq{\bvec{s}^u}; \bmat{W} \right),
\end{equation}
in which $\mathcal{R}(\cdot\;; \bmat{W})$ represents the function the RNN computes with parameters $\bmat{W}$.

To learn a semantically rich user taste vector that is able to generate adequate recommendations, ideally this taste vector should be able to capture how a user's listening behavior is evolving over time.
We therefore train the RNN to predict a song the user is going to listen to in the future.
More specifically, for a particular user $u$, we take the first $n$ consecutive songs $\bvec{s}^u_{1:n}$ this user has listened to, and we try to predict a future song vector $\bvec{s}^u_{n+\ell}$, for some strictly positive value of $\ell$.
As a loss function, we use the $L_2$ distance between the predicted taste vector and the true future song vector:
\begin{align}
\loss{\bvec{s}^u_{n+\ell}, \mathcal{R}\left( \bvec{s}^u_{1:n}; \bmat{W} \right)} = \norm{\bvec{s}^u_{n+\ell} - \mathcal{R}\left( \bvec{s}^u_{1:n}; \bmat{W} \right)}_2.
\end{align}
We experimented with other distance functions, such as cosine distance and a weighted mixture of cosine and $L_2$ distance.
The cosine distance $L_{\cos}(\cdot)$ between two vectors $\vec{x}$ and $\vec{y}$ is given by:
\begin{align}
L_{\cos}\left(\vec{x}, \vec{y}\right) = 1 - \frac{\vec{x}\cdot \vec{y}}{\norm{\vec{x}}_2 \norm{\vec{y}}_2}.
\end{align}
The $L_2$ distance, nevertheless, gave the best results in the experiments.

To determine the best value of the prediction offset $\ell$ we consider two separate prediction tasks: short-term and long-term prediction.
The idea is that it is generally easier to predict a song a user is going to listen to next---e.g.~think of a user shuffling an artist album, or listening to a rock playlist---than it is to predict the 50th song he is going to listen to in the future.
The underlying dynamics of a short-term and long-term prediction model will therefore be different.
For example, the short-term model will intuitively be more focused on the last few tracks that were played, while the long-term model will generally look at a bigger timeframe.
During training we sample a value of $\ell$ for every input sequence from a discrete uniform distribution.
More specifically, for the short-term model, $\ell$ is sampled from $\uniform{1}{10}$, and $\ell$ is sampled from $\uniform{25}{50}$ for the long-term model.
Random sampling of the prediction offset for every new input sequence reduces chances of overfitting and also increases model generalization.
The entire training procedure is sketched in Algorithm \ref{algo:training}. In here, we use a stochastic minibatch version of the gradient-based Adam algorithm to update the RNN weights \cite{Kingma:2015ku}.
We have also experimented with setting $\ell$ fixed to 1 for the short-term model, but this leads to very near-sighted models that make predictions almost solely based on the last played song.
For example, first listening to 100 classical songs and then to 1 pop song would lead to pop song predictions by the RNN.

\begin{algorithm}[t!]
\DontPrintSemicolon
	\SetKwFunction{Adam}{adam\_update}
	\SetKwFunction{Shuffle}{shuffle}
	\SetKwData{Loss}{loss}
	\SetKwInOut{Input}{input}
	\SetKwInOut{Output}{output}
	\SetKwInOut{Parameter}{parameter}
	\Input{dataset $\mathcal{D}$ of song sequences, initial RNN parameters $\bmat{W}$}
	\Parameter{sequence length $n$, offsets $\ell_{\min}$ and $\ell_{\max}$, learning rate $\eta$}
	\Repeat{convergence}{
		\Shuffle{$\mathcal{D}$}\;
		\ForEach{batch $B \in \mathcal{D}$}{
		\Loss $\leftarrow 0$\;
			\ForEach{sequence $(\bvec{s}) \in B$}{
				$\ell \sim \uniform{\ell_{\min}}{\ell_{\max}}$\;
				\Loss $\leftarrow$ \Loss $+ \;\loss{\bvec{s}_{n+\ell}, \mathcal{R}\left( \bvec{s}_{1:n}; \bmat{W} \right)}$\;
			}
			$\bmat{W} \leftarrow$ \Adam{$\bmat{W}$, \Loss, $\eta$}\;
		}
	}
  \caption{RNN training procedure}\label{algo:training}
\end{algorithm}

\subsection{Recommending songs}
\label{subsec:recommending}
The output of the RNN is a user taste vector that resides in the same vector space as all songs in the catalog.
Since we trained the RNN to produce these taste vectors to lie close to future songs a user might play---in terms of $L_2$ distance---we can query for nearby songs in that vector space to generate new song recommendations.
To scale the search for nearby songs in practice, we construct an Annoy\footnote{\url{github.com/spotify/annoy}} tree datastructure with all song vectors.
The Annoy tree will iteratively divide the vector space in regions using a locality-sensitive hashing random projection technique \cite{Charikar:2002km}, which facilitates approximate nearest neighbor querying.
To generate suggested recommendations for a particular user, we query the Annoy tree using the user's taste vector to find the top $k$ closest catalog songs in the vector space.

\subsection{Incorporating play context}
\label{sec:playcontext}
In general we do not only know the order in which a user plays particular songs, but we also know the context in which the songs have been played.
By context we mean a playlist, an album page, whether the user deliberately clicked on the song, etc.
This additional information can be very useful, e.g.~we can imagine that a user clicking on a song is a stronger indicator of the user's taste than when the song is played automatically in an album or playlist after the previous song has finished playing.
Since the RNN can process any combination of arbitrary embeddings and features, we can supply a context vector $\bvec{c}^u$ as extra input.
The context vector is in this case concatenated with the song vector at each time step to produce the user taste vector:
\begin{equation}
\forall u \in U\colon \bvec{t}^u = \mathcal{R}\left( \seq{\bvec{s}^u \oplus \bvec{c}^u}; \bmat{W} \right),
\end{equation}
in which we use the symbol $\oplus$ to denote vector concatenation.
To construct a context vector we consider the ordered set $C = (\text{album}, \text{playlist}, \text{artist}, \text{click}, \dots)$ of all possible contexts, denote $C_i$ as the $i$'th context in $C$, and $c(s)$ as the play context for a particular song $s$, e.g.~$c(s) = \set{\text{playlist}, \text{click}}$.
The context vector for a song $s$ is then constructed using a one-hot encoding scheme:
\begin{align}
\bvec{c} = \sum_{i=1}^{|C|} \mathrm{onehot}(i, |C|) \cdot \mathbf{1}_{c(s)}(C_i),
\end{align}
in which $\mathrm{onehot}(i, |C|)$ is a one-hot vector of length $|C|$ with a single $1$ at position $i$, and in which $\mathbf{1}_{A}(x)$ is the indicator function that evaluates to $1$ if $x\in A$ and to $0$ otherwise.
We also include the time difference between playing the current song and the last played song.
The final context vector $\bvec{c}^u_j$ for the $j$'th song $s^u_j$ played by user $u$ then becomes:
\begin{align}
\bvec{c}^u_j = \Delta(s^u_j, s^u_{j-1}) \oplus \sum_{i=1}^{|C|} \mathrm{onehot}(i, |C|) \cdot \mathbf{1}_{c\left(s^u_j\right)}(C_i),
\end{align}
in which $\Delta(s_j, s_{j-1})$ is the time difference in seconds between playing song $s_j$ and $s_{j-1}$, evaluating to $0$ if $s_{j-1}$ does not exist.

\subsection{User and model updates}
\label{sec:update}
The recommendation pipeline we described in this section can be used in both static and dynamic contexts.
In dynamic contexts, the model should be updated frequently so that recommendations can immediately reflect the user's current listening behavior.
This is easily done in our model: for every song the user listens to we retrieve its song vector that we use to update the hidden state of the RNN, after which we calculate a new taste vector to generate recommendations.
This requires that we keep track of the current RNN states for every user in the system, which is a small overhead.
In more static contexts, in which recommendations are generated on a regular basis for all users (every day, week\dots), there is no need to update the RNN for every song a user listens to.
Here, we retrieve the entire user listening history--or the last $n$ songs--which we feed to a newly initialized RNN.
We therefore do not need to remember the RNN states for every user.

All recommendation modules that are deployed in practical environments have to be updated regularly in order to reflect changes in the item catalog and user behavior.
This is no different for the framework we present here.
In order to perform a full model update, we subsequently train word2vec on the playlists in the catalog, retrain the RNN model on the new song vectors, and populate the Annoy tree with the same song vectors.
Only retraining word2vec is not sufficient since we almost never end up in the same song embedding space.
If we use dynamic user updates, note that we also have to do one static update after retraining word2vec and the RNN, since the remembered RNN states will have become invalid.

\section{Data Gathering and Analysis}
\label{sec:datagathering}
The first stage in the recommendation pipeline considers the learning of semantically rich song vectors, for which we use the word2vec algorithm.
In this section we explain how we gather data to train this word2vec model, and we perform a preliminary analysis on the song vectors.
Finally we detail the construction of the train and test data for the RNNs.

\subsection{Training word2vec}
To create training data for the word2vec model we treat user-created playlists as documents and songs within these playlists as individual words, as mentioned in Section \ref{sec:methodology}.
For this, we gather all user-created playlists on the Spotify music streaming platform.
In these playlists we only consider the top $N$ most popular tracks; the other tracks are removed.
In our experiments we set $N$ to 6 million, which makes up for most of the streams on the Spotify platform.
After filtering out unpopular tracks, we further only consider playlists with a length larger than 10 and smaller than 5000.
We also restrict ourselves to playlists which contain songs from at least 3 different artists and 3 different albums, to make sure there is enough variation and diversity in the playlists.
After applying the filtering above, we arrive at a corpus of $276.5$ million playlists in total.
In the following step, the playlists are fed to the word2vec suite, where we use the CBoW algorithm with negative sampling.
We go through the entire playlist corpus once during execution of the algorithm in order to produce vectors with dimensionality $D=40$, a number that we empirically determined and produces good results.

\subsection{Data processing and filtering}
\label{sec:datafiltering}
In Section \ref{sec:experiments} we will train different RNN versions, for which we will use both playlist and user listening data.
Playlists are usually more contained regarding artists and musical genres compared to listening data.
Modeling playlist song sequences with RNNs will therefore be easier, but we can miss out important patterns that appear in actual listening data, especially if a user listens to a wide variety of genres.

For the playlist data we extract chunks of 110 consecutive songs, i.e.~60 songs to feed to the RNN and the next 50 songs are used as ground truth for the prediction.
Regarding the user listening, which is captured in the first half of 2016 on the Spotify platform, we only keep songs which have a song vector associated with them; other songs are removed from the user's history.
We also save context information for every played song; for this, we consider the following 13 Spotify-related contexts: \textit{collection}, \textit{library}, \textit{radio}, \textit{own playlist}, \textit{shared playlist}, \textit{curated playlist}, \textit{search}, \textit{browse}, \textit{artist}, \textit{album}, \textit{chart}, \textit{track}, \textit{clicked}, and \textit{unknown} for missing context information.
To extract RNN training sequences we take chunks of 150 consecutive songs, for which the first 100 songs are again used as input to the RNN and the last 50 as ground truth.
We allow more songs as input to the RNN compared to the playlist training data since user listening data is generally more diverse than playlist data, as mentioned before.

Since users often return to the same songs or artists over and over again, we apply additional filtering to eliminate those songs.
This will greatly improve the RNN's generalization, and will counter what we call the `easy prediction bias', as it is too easy to predict songs a user has already listened to.
This filtering is not needed for playlist data, since a song only appears once in a playlist most of the times.
The filtering rules we include, are:
\begin{enumerate}
\item The last 50 songs should be unique;
\item The last 50 songs should not appear in the first 100 songs;
\item The last 50 artists should be unique;
\item The last 50 artists should not appear in the first 100 artists.
\end{enumerate}
Note that we only remove similar songs and artists from the ground truth labels in the dataset, and that we leave the first 100 songs in the user's history intact.
That is, the same song can still appear multiple times in these first 100 songs, thereby steering the user's preference towards this particular song and influencing the predictions made by the RNN.
For both playlist and listening data we gather 300,000 train sequences, 5,000 validation sequences and 5,000 test sequences.

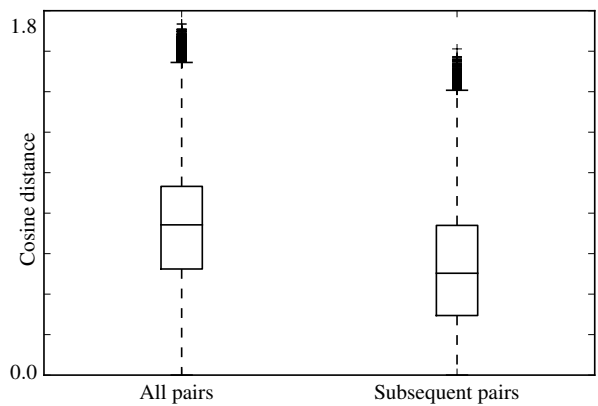
\begin{figure}[t!]
\begin{tikzpicture}[scale=.90,inner sep=0pt]
	\node [draw=none, anchor=south west] () at (0,0) {\includegraphics[trim = 1.9cm 1.2cm 1.5cm 1cm, clip, width=.44\linewidth]{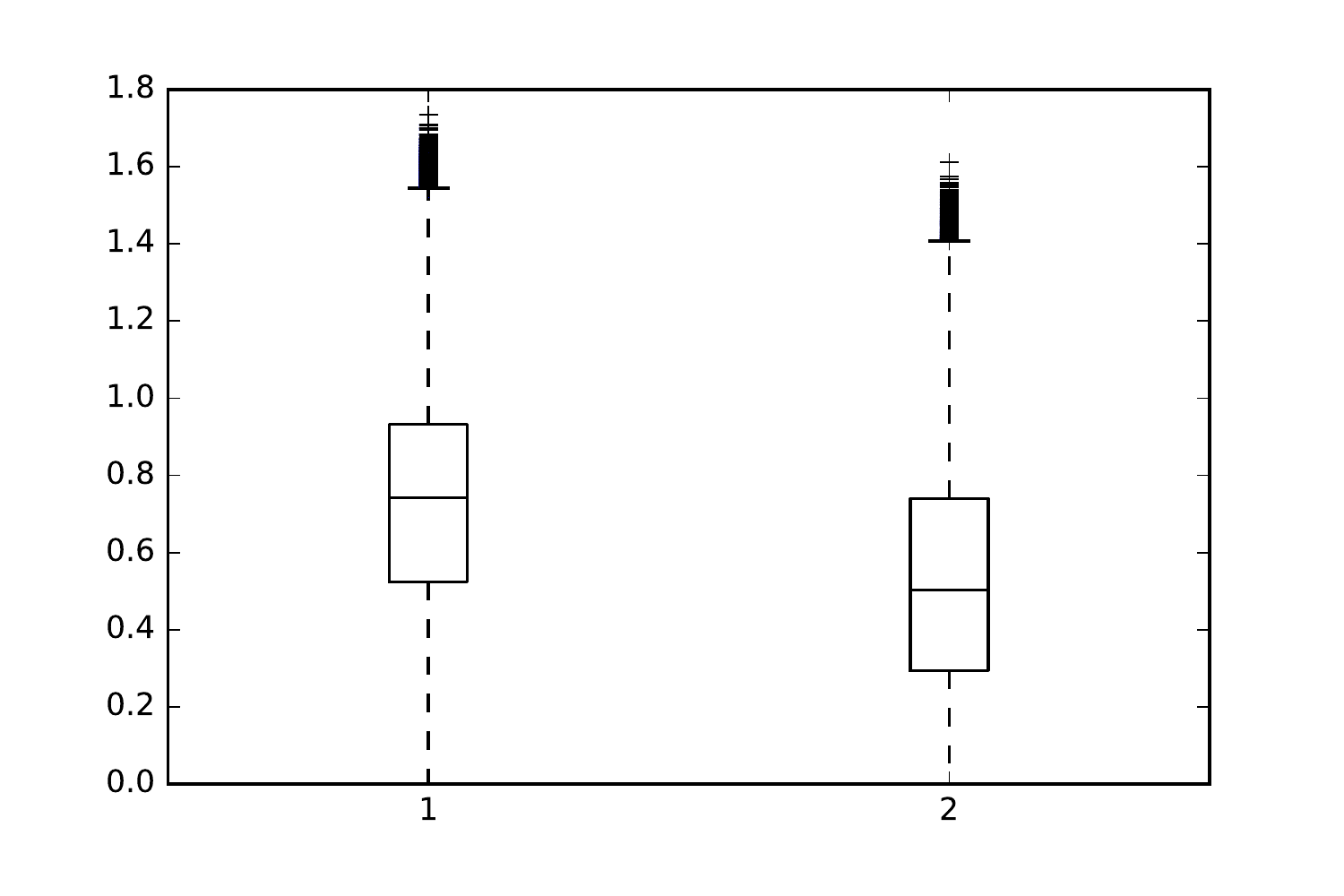}};

	\node[below=2pt] (b1) at (1.95,0.0) {All pairs};
	\node[below=2pt] (b3) at (5.9,0.0) {Subsequent pairs};
	\draw[draw opacity=0] (0.0, 0.1) -- node[left=3pt] {\rotatebox{90}{Cosine distance}} (0.0, 5.2);
	\node[left=2pt] (lO) at (0.0,0.1) {$0.0$};
	\node[left=2pt] (l1) at (0.0,5.2) {$1.8$};
\end{tikzpicture}
\caption{Box plots of pairwise distances between songs in all test set listening histories.}
\label{fig:boxplots}
\end{figure}

\begin{figure}[t!]
\begin{tikzpicture}[scale=.90,inner sep=0pt]
	\node [draw=none, anchor=south west] () at (0,0) {\includegraphics[trim = 1.9cm 1.2cm 1.5cm 1cm, clip, width=.44\linewidth]{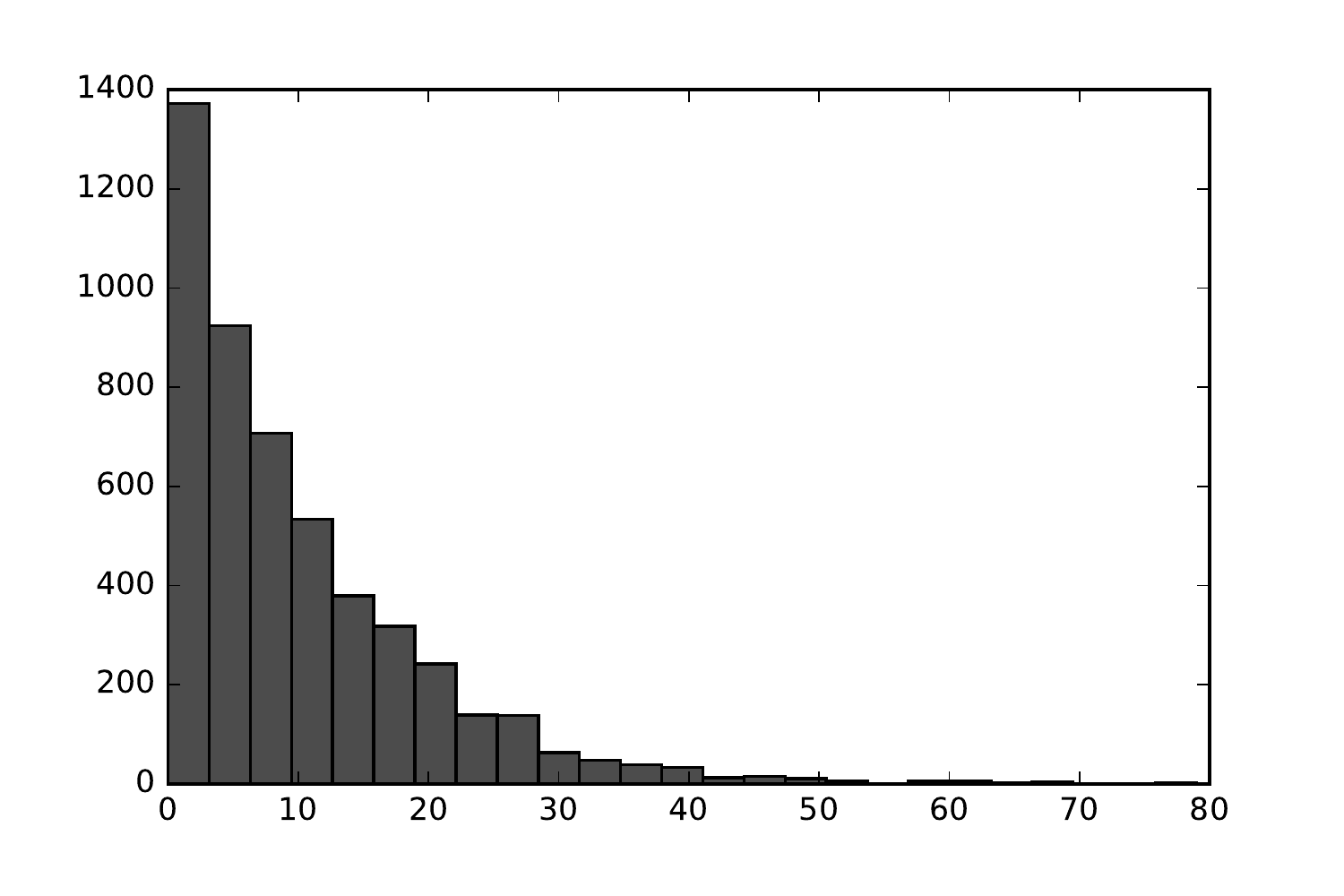}};

		\node[below=2pt] (bO) at (0.1,0.0) {$0$};
	\node[below=2pt] (b1) at (1.0,0.0) {$10$};
	\node[below=2pt] (b2) at (1.97,0.0) {$20$};
	\node[below=2pt] (b3) at (2.94,0.0) {$30$};
	\node[below=2pt] (b4) at (3.91,0.0) {$40$};
	\node[below=2pt] (b4) at (4.88,0.0) {$50$};
	\node[below=2pt] (b4) at (5.85,0.0) {$60$};
	\node[below=2pt] (b4) at (6.82,0.0) {$70$};
	\node[below=2pt] (b4) at (7.79,0.0) {$80$};
	\draw[draw opacity=0] (0.1,0.0) -- node[below=12pt] {$\#$transitions with cos distance $> 1$} (7.79,0.0);
	\draw[draw opacity=0] (0.0, 0.0) -- node[left=20pt] {\rotatebox{90}{$\#$songs}} (0.0, 5.2);
	
	\node[left=2pt] (lO) at (0.0,0.1) {$0$};
	\node[left=2pt] (lO) at (0.0,1.12) {$200$};
	\node[left=2pt] (lO) at (0.0,2.14) {$400$};
	\node[left=2pt] (lO) at (0.0,3.16) {$600$};
	\node[left=2pt] (lO) at (0.0,4.18) {$800$};
	\node[left=2pt] (l1) at (0.0,5.2) {$1000$};
\end{tikzpicture}
\caption{Histogram of the number of song transitions with cosine distance larger than 1.}
\label{fig:histogram-cos-gte-1}
\end{figure}
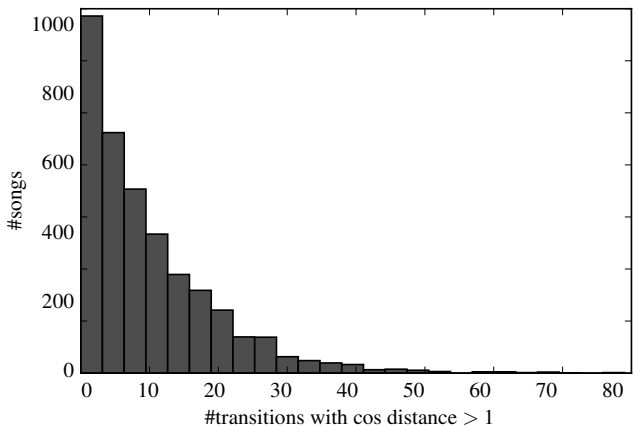

\subsection{User data analysis}
To analyze the gathered data, we take all 5,000 listening history sequences in the test set, and we calculate pairwise cosine distances between the song vectors in these sequences.
We measure both the vector distance between all possible song pairs in a listening sequence, as well as the distance only between subsequent songs.
The distances are given as box plots in Figure \ref{fig:boxplots}, in which the whiskers are drawn at 1.5 times the interquartile range.
We see that the all pairs median distance is larger than the subsequent pairs median distance by around 0.25, indeed confirming the higher correlation between subsequent songs.
We also plot the number of song transitions within each user's listening history that have a cosine distance larger than 1, meaning that the next song is more different from the current song than it is similar to it.
The histogram for this is shown in Figure \ref{fig:histogram-cos-gte-1}.
Most listening histories have little such song transitions.
The median number is seven, which points, on average, to listening periods of 21 similar songs before transitioning to a more different region in the song space, which in turn corresponds to coherent listening sessions of about 1 to 1.5 hours long.

\section{Experiments}
\label{sec:experiments}
In this section we discuss the results of various experiments and we show examples of song recommendations produced by different RNN models.
In the following we first design the best RNN architecture by experimentation, after which we explain the baselines against which the RNN models will be tested.
All experiments are run on an Amazon AWS instance, 32-core Intel Xeon 2.60GHz CPU, 64GB RAM and Nvidia GRID K520 GPU with cuDNN v4.
The RNN models are implemented in Lasagne\footnote{\url{github.com/Lasagne/Lasagne}} and Theano \cite{AlRfou:2016uc}.

\subsection{Network architecture}
\label{sec:architecture}
The neural network architecture we propose consists of a number of recurrent layers followed by a number of dense layers.
The number of input dimensions is $D = 40$, which is the dimensionality of a single song vector.
For the recurrent layers we consider both LSTMs and GRUs, since such models are state-of-the-art in various NLP applications and can handle long-term dependencies.
We use standard implementations of these layers as described in Section \ref{sec:motivation}, with a hidden state dimensionality of all layers equal to $D_{hid}$.
We will describe how the optimal number of recurrent layers and the optimal value of $D_{hid}$ are chosen.

After the recurrent layers we consider two dense layers.
We empirically set the dimensionality of the first dense layer to $4\cdot D_{hid}$, and we use a leaky ReLU nonlinearity with a leakiness of 0.01 \cite{Maas:2013tn}.
Since we are predicting a user taste vector in the same song space, the output dimensionality of the last dense layer is $D = 40$, which is the same as the input dimensionality.
In the last layer we use a linear activation function, since we are predicting raw vector values.

In a first experiment we set $D_{hid}$ to 50, we switch the recurrent layer type between LSTM and GRU, and we also vary the number of recurrent layers from 1 up to 3.
We record train loss, validation loss, average gradient update time for a single batch and average prediction time for a single sequence.
For this experiment we use the gathered playlist data, and we train a short-term model for which we report the best values across 15 epochs of training.
The results are shown in Table \ref{table:compare-rnn-types}.
We see that both models are very comparable in terms of validation loss, and that the optimal number of recurrent layers is two.
The LSTM model performs slightly better than the GRU model in this case, but the GRU model predicts more than 50\% faster than the LSTM model, which can be important in large-scale commercial production environments.
We observe comparable results for listening history prediction as well as long-term prediction.
We therefore pick the two-layer GRU model as the optimal architecture.

\begin{table}[t!]
\small
\caption{Comparing recurrent layer type and number of layers on short-term playlist prediction.}
\label{table:compare-rnn-types}
\tabcolsep=0.07cm
\begin{tabular}{l | c c c c}
\toprule
				& Train loss & Validation loss & Update [ms] & Prediction [ms]\\
\hline
LSTM, 1 layer		& 258.8	& 271.2	& 20		& 	 4.4	\\
LSTM, 2 layers		& 257.3	& 270.2	& 39		& 	 8.7	\\
LSTM, 3 layers		& 258.1	& 271.0	& 58		& 	13.0	\\
\hline
GRU, 1 layer			& 259.0	& 271.0	& 16		& 	2.2	\\
GRU, 2 layers		& 257.3	& 270.5	& 30		& 	4.1	\\
GRU, 3 layers		& 257.5	& 270.6	& 45		& 	6.1	\\
\bottomrule
\end{tabular}
\end{table}

\begin{table}[t!]
\small
\caption{Comparing performance with varying hidden layer size $D_{hid}$ on short-term playlist prediction.}
\label{table:compare-hidden-sizes}
\tabcolsep=0.15cm
\begin{tabular}{c | c c}
\toprule
$D_{hid}$ & $L_2$ validation loss & Cosine validation loss\\
\hline
20		  & 287.3	& 0.464\\
30		  & 276.4	& 0.425\\
40		  & 271.5	& 0.408\\
50		  & 270.5	& 0.406\\
60		  & 270.0	& 0.406\\
70		  & 270.4	& 0.406\\
80		  & 270.5	& 0.407\\
90		  & 270.1	& 0.405\\
100		  & 270.4	& 0.406\\
\bottomrule
\end{tabular}
\end{table}

Next we perform experiments to determine the optimal hidden state size $D_{hid}$.
We train the two-layer GRU architecture from above for a short-term playlist prediction, and we vary $D_{hid}$ from 20 to 100 in steps of 10.
The results are shown in Table \ref{table:compare-hidden-sizes}.
The validation loss is the highest at 20, and is minimal at $D_{hid}$ values of 50 and 60.
The loss remains more or less stable if we increasing the hidden state size to 100.
Since larger hidden state sizes imply slower prediction and train times, we choose $D_{hid} = 50$ as the optimal hidden state size.
These observations are also valid for long-term prediction and for listening history data.
The final architecture is displayed in Table \ref{table:architecture}.
It turns out that this architecture is also near-optimal for long-term playlist prediction as well as for user listening data.
Since the number of network parameters is low and given the large amount of training data, we do not need additional regularization.

We also train short-term user listening RNNs with additional play context information at the input, as described in Section \ref{sec:playcontext}, and with the same configurations as in Tables \ref{table:compare-rnn-types} and \ref{table:compare-hidden-sizes}.
The optimal configuration is a 2-layer architecture with $D_{hid}$ equal to 100, but the differences in performance for $50 \leq D_{hid} < 100$ are minimal.
Furthermore, the performance gain compared to the same model without context information is non-significant.
We will therefore disregard play context models in the rest of the experiments.

\subsection{Baselines}
\label{sec:baselines}
We will compare the performance of different RNN models against several baseline models.
The general idea here is again that a user's listening behavior is governed by underlying temporal dynamics.
We consider three types of baseline models: an exponential discount model, a weight-based model, and a classification model.
The first two types are aggregation models, which means that they take an arbitrary number of song vectors as input and produce one output vector by mathematically combining the song vectors, e.g.~through summing or taking a maximum across dimensions.
Aggregation of distributed vectors is a popular practice in NLP applications and deep learning in general since it does not require any training when the vector space changes \cite{dosSantos:2014tr, Collobert:2011tk}.
In our case however, the danger of aggregation is that sometimes songs from different genres are summed together, so that we can end up in `wrong' parts of the song space.
For example, if we aggregate the song vectors of a user listening to a mix of classical and pop music, we might arrive in a space where the majority of songs is jazz, which in turn will lead to bad recommendations.
The third type of baseline is based on the work by Hidasi et al.~\cite{Hidasi:2015uq} as mentioned in Section \ref{sec:motivation}.
In this, we will not predict an output vector to query the Annoy LSH tree, but we will output probability scores for all $N$ items in the catalog. The top $k$ items can then immediately be recommended.

\begin{table}[t!]
\small
\caption{The final neural network architecture.}
\label{table:architecture}
\tabcolsep=0.15cm
\begin{tabular}{c | l}
\toprule
          & Layer type (no.~of dimensions) and non-linearity\\
\hline
		  & Input (40)\\
1		  & GRU (50)\\
		  & \qquad sigmoid (gates); tanh (hidden update)\\
2		  & GRU (50)\\
          & \qquad sigmoid (gates); tanh (hidden update)\\
3         & Fully connected dense (200)\\
          & \qquad Leaky ReLU, leakiness $= 0.01$\\
4         & Fully connected dense (40)\\
          & \qquad Linear activation\\
\bottomrule
\end{tabular}
\end{table}

\subsubsection*{Exponential discount model}
In the exponential discount model we make the assumption that a user's taste is better reflected by the recent songs he has listened to than by songs he has listened to a while ago.
We model this type of temporal dynamics by the following exponentially decaying weight model:
\begin{equation}
\forall u \in U\colon \bvec{t}^u = \sum_{j=1}^k \bvec{s}^u_j \cdot \gamma^{k-j}.
\end{equation}
In this, we consider the song history of user $u$ which has length $k$, and we weigh every vector by a power of $\gamma$, the discount factor.
Setting $\gamma = 1$ results in no discounting and leads to a simple sum of the song vectors, while setting $\gamma < 1$ focuses more attention on recently played songs.
The smaller $\gamma$, the more substantial this contribution becomes compared to songs played a longer time ago.
If $\gamma = 0$, the user taste vector is essentially the vector of the last played song.

\subsubsection*{Weight-based model}
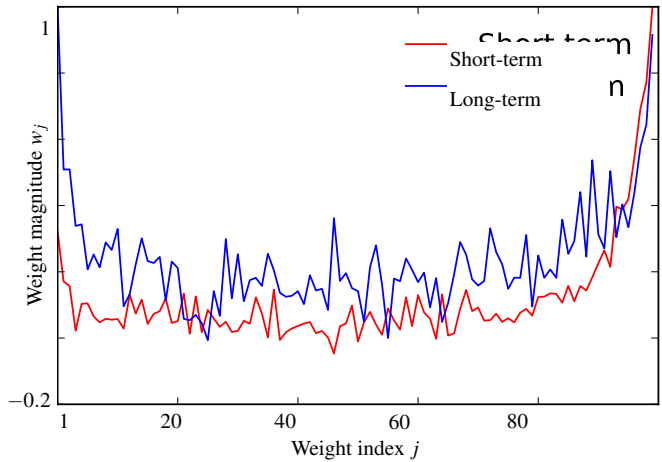
\begin{figure}[t!]
\begin{tikzpicture}[scale=.97,inner sep=0pt]
	\node [draw=none, anchor=south west] () at (0,0) {\includegraphics[trim = 1.9cm 1.2cm 1.5cm 1cm, clip, width=.48\linewidth]{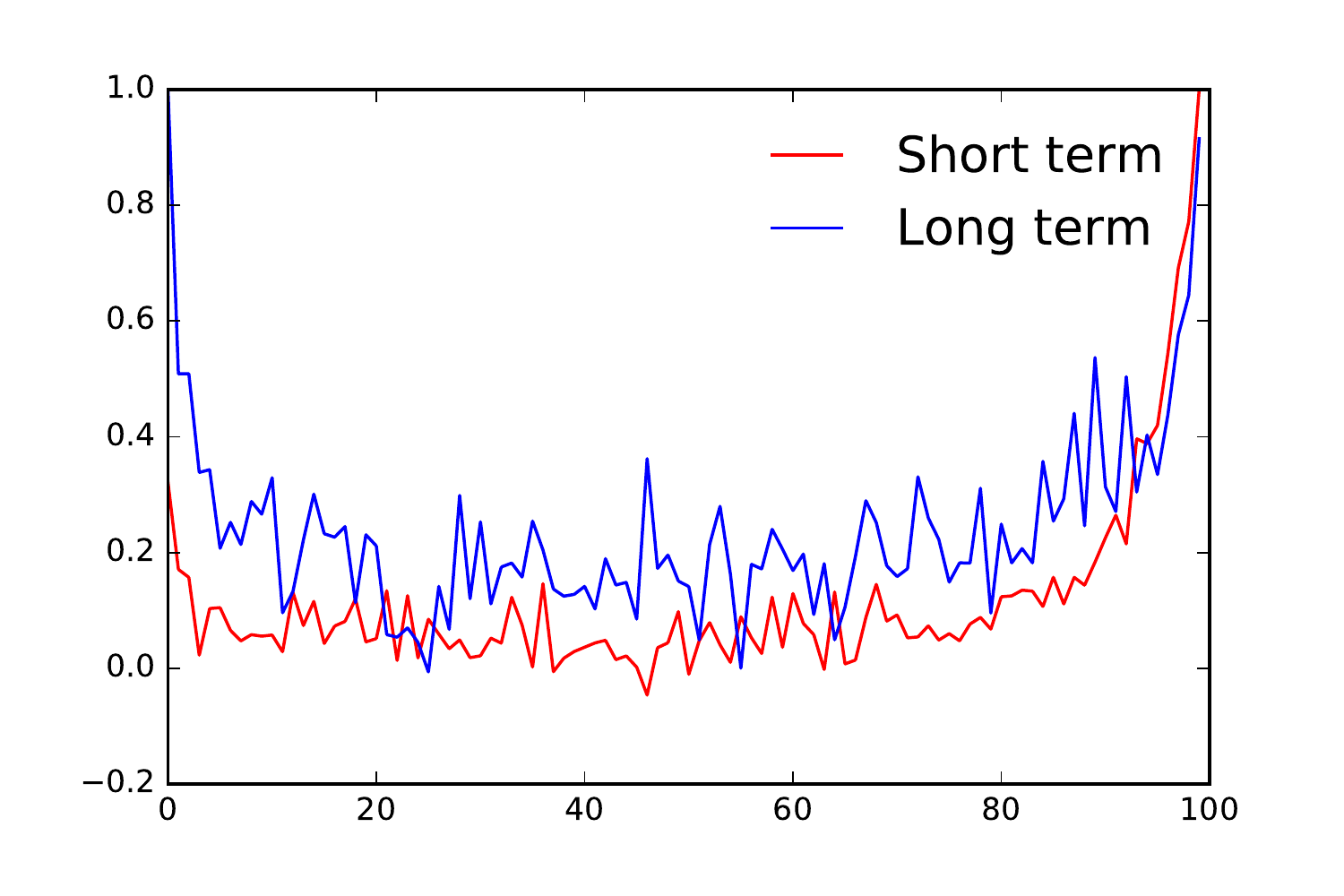}};

	\node[below=2pt] (bO) at (0.1,0.0) {$1$};
	\node[below=2pt] (b1) at (1.55,0.0) {$20$};
	\node[below=2pt] (b2) at (3.1,0.0) {${40}$};
	\node[below=2pt] (b3) at (4.65,0.0) {${60}$};
	\node[below=2pt] (b3) at (6.25,0.0) {${80}$};
	\draw[draw opacity=0] (0.1,0.0) -- node[below=12pt] {Weight index $j$} (8.0,0.0);
	\draw[draw opacity=0] (0.0, 0.1) -- node[left=3pt] {\rotatebox{90}{Weight magnitude $w_j$}} (0.0, 5.2);
	\node[left=2pt] (lO) at (0.0,0.1) {$-0.2$};
	\node[left=2pt] (l1) at (0.0,5.2) {$1$};
	
	\fill[white] (5.4,4) rectangle (7.5,5);
	\node[right=2pt] (t1) at (5.25,4.78) {Short-term};
	\node[right=2pt] (t2) at (5.25,4.2) {Long-term};
\end{tikzpicture}
\caption{Normalized weight magnitudes for long- and short-term prediction.}
\label{fig:weight_plot}
\end{figure}

This model is based on the weighted word embedding aggregation technique by De Boom et al.~\cite{DeBoom:2016gm}.
As in the exponential discount model, we will multiply each song vector $\bvec{s}_j$ by a weight $w_j$:
\begin{equation}
\forall u \in U\colon \bvec{t}^u = \mathcal{W}\left( \seq{\bvec{s}^u}; \bvec{w} \right) = \sum_{j=1}^k \bvec{s}^u_j \cdot w_j,
\end{equation}
in which we gather all weights $w_j$ in a $k$-dimensional vector $\bvec{w}$.
Now, instead of fixing the weights to an exponential regime, we will learn the weights through the same gradient descent procedure as in Algorithm \ref{algo:training}.
In this algorithm we replace the RNN loss by the following weight-based loss:
\begin{align}
&\loss{\bvec{s}^u_{k+\ell}, \mathcal{W}\left( \seq{\bvec{s}^u}; \bvec{w} \right)} \nonumber\\
&\qquad\qquad\qquad\quad= \norm{\bvec{s}^u_{k+\ell} - \mathcal{W}\left( \seq{\bvec{s}^u}; \bvec{w} \right)}_2 + \lambda\norm{\bvec{w}}_2.
\end{align}
We include the last term as weight regularization, and we empirically set $\lambda$ to 0.001.
Apart from this regularization term, we do not imply any restrictions on or relations between the weights.
We train a weight-based model on user listening data for both short- and long-term predictions, and the resulting weights are plotted in Figure \ref{fig:weight_plot}.
For the short-term prediction the weights are largest for more recent tracks and decrease the further we go back in the past.
This confirms the hypothesis that current listening behavior is largely dependent on recent listening history.
For the long-term model we largely observe the same trend, but the weights are noisier and generally larger than the short-term weights.
Also, the weights increase again in magnitude for the first 10 tracks in the sequence.
This may signify that a portion of a user's long-term listening behavior can be explained by looking at songs, genres or artists he has listened to in the past and returns to after a while.

\subsubsection*{Classification model}
As mentioned above, we loosely rely on the work by Hidasi et al.~\cite{Hidasi:2015uq} to construct a classification baseline model.
In this work, the items are encoded as a one-hot representation, and the output of the model is a probability score for every item in the catalog.
To be able to fairly compare between all other baselines, and to help scale the model, we slightly adapt it and use the word2vec vectors as input instead of the one-hot item encodings.
We employ the same neural network model as in Table \ref{table:architecture} and at the output we add an additional dense layer with output dimensionality $N=6{,}000{,}000$ and softmax activation function.
The memory footprint of this softmax model thus substantially increases by around 938MiB, compared to the model in Table  \ref{table:architecture}.
Given the large output dimensionality, we also experimented with a two-stage hierarchical softmax layer \cite{Goodman:2001dp}, but the computational improvements were only marginal and the model performed worse.

We train the softmax classification model with two different loss functions.
First, we consider the categorical cross-entropy loss in the case there is only one target:
\begin{align}
\loss{i, \mathcal{R}\left( \bvec{s}^u_{1:k}; \bmat{W} \right)} = - \mathrm{onehot}\left(s^u_{k+\ell}, N\right) \cdot \log \left[ \mathrm{softmax}\left(\mathcal{R}\left( \bvec{s}^u_{1:k}; \bmat{W} \right) \right) \right]^\top.
\end{align}
In this loss function, $i$ is the RNN output index of the target song to be predicted, and $\mathrm{softmax}\left(\mathcal{R}\left( \bvec{s}^u_{1:k}; \bmat{W} \right)\right)$ is the output of the RNN after a softmax nonlinearity given the input vectors $\bvec{s}^u_{1:k}$.
The second loss function is a pairwise ranking loss used in the Bayesian Personalized Ranking (BPR) scheme by Rendle et al.\cite{Rendle:2009wp}.
This loss function evaluates the score of the positive target against randomly sampled negative targets:
\begin{align}
\loss{i, \mathcal{R}\left( \bvec{s}^u_{1:k}; \bmat{W} \right)} = - \frac{1}{N_S} \sum_{j=1}^{N_S} \log\left[ \sigma\left( \mathcal{R}\left( \bvec{s}^u_{1:k}; \bmat{W} \right)_i - \mathcal{R}\left( \bvec{s}^u_{1:k}; \bmat{W} \right)_j \right) \right].
\end{align}
In this, $N_S$ is the number of negative samples that we set fixed to 100, $\sigma(\cdot)$ is the sigmoid function, and $i$ is again the output index of the positive sample.
Note that we use a sigmoid nonlinearity rather than a softmax.
In practice we also add an $L2$ regularization term on the sum of the positive output value and negative sample values.
To generate negative samples, we sample song IDs from a Zeta or Zipf distribution with parameter $z = 1.05$, which we checked empirically on the song unigram distribution:
\begin{align}
\mathrm{Zipf}_z(k) = \frac{k^{-z}}{\zeta(z)},
\end{align}
in which $\zeta(\cdot)$ is the Riemann zeta function.
We resample whenever a song appears in a user's listening data to make sure the sample is truly negative.

Hidasi et al.~reported better stability using BPR loss compared to cross-entropy loss, but our sampling-based training procedure from Algorithm \ref{algo:training} did not produce any unstable networks for both loss functions.
We trained short-term and long-term networks on the filtered listening history data using both cross-entropy and BPR, and all models took around 2.5 days to converge.
By comparison, training until convergence on the same hardware only took 1.5 hours for the models presented in this work.

%

\subsection{Results}
\label{sec:results}
In this section we will display several performance metrics on the test set of user listening histories.
After all, the music recommendations will be based on what a user has listened to in the past.
We have trained four RNN models: a playlist short-term (\textit{rPST}) and long-term (\textit{rPLT}) RNN, and a user history short-term (\textit{rHST}) and long-term (\textit{rHLT}) RNN.
We also report metrics for five baselines: a short-term (\textit{bWST}) and long-term (\textit{bWLT}) weight-based model, and exponential discount models with $\gamma \in \set{1.0, 0.97, 0.85}$.
In the following we will perform a forward analysis to evaluate how well a taste vector is related to future song vectors, which will show the predictive capacity of the different models.
We will also do a backwards analysis to study on what part of the listening history sequences the different models tend to focus.
We conclude with results on a song prediction task.

\subsubsection*{Forward analysis}
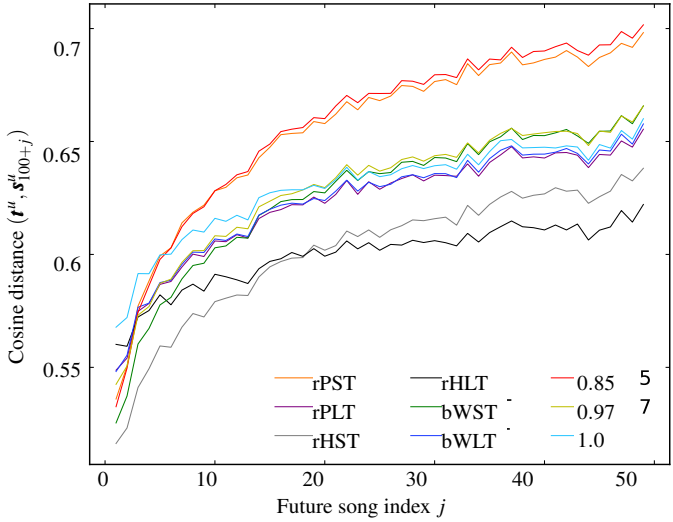
\begin{figure}[t!]
\begin{tikzpicture}[scale=.85,inner sep=0pt]
	\node [draw=none, anchor=south west] () at (0,0) {\includegraphics[trim = 3.1cm 2.5cm 2.5cm 2.0cm, clip, width=.465\linewidth]{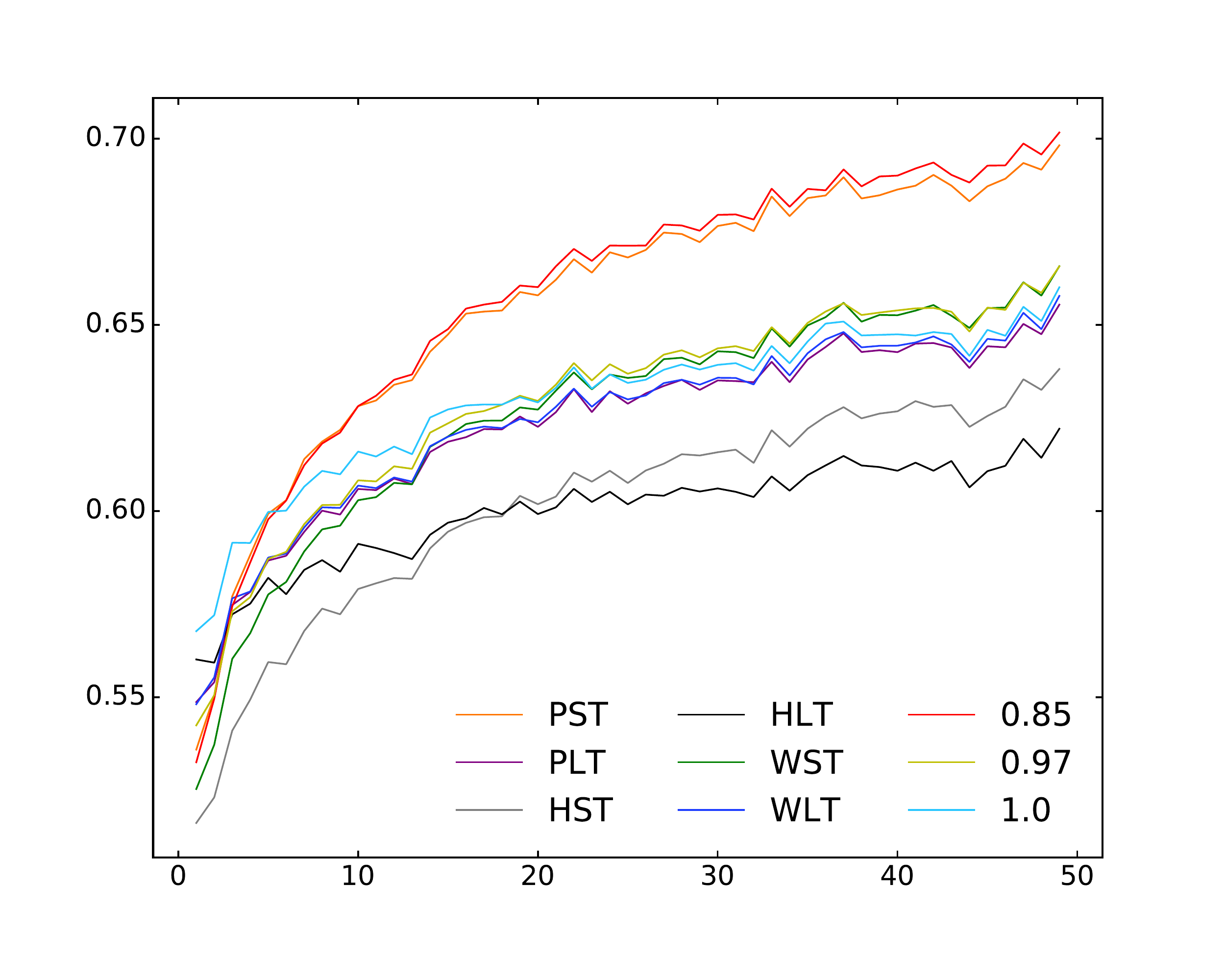}};

	\node[below=2pt] (bO) at (0.25,0.0) {$0$};
	\node[below=2pt] (b1) at (1.85,0.0) {$10$};
	\node[below=2pt] (b2) at (3.5,0.0) {${20}$};
	\node[below=2pt] (b3) at (5.1,0.0) {${30}$};
	\node[below=2pt] (b4) at (6.75,0.0) {${40}$};
	\node[below=2pt] (b5) at (8.35,0.0) {${50}$};
	\draw[draw opacity=0] (0.1,0.0) -- node[below=12pt] {Future song index $j$} (8.4,0.0);
	\draw[draw opacity=0] (0.0, 0.1) -- node[left=20pt] {\rotatebox{90}{Cosine distance $(\bvec{t}^u, \bvec{s}^u_{100+j})$}} (0.0, 6.75);
	\node[left=2pt] (lO) at (0.0,1.5) {$0.55$};
	\node[left=2pt] (l1) at (0.0,3.15) {$0.6$};
	\node[left=2pt] (l2) at (0.0,4.85) {$0.65$};
	\node[left=2pt] (l3) at (0.0,6.5) {$0.7$};
	
	\fill[white] (3.5,0.2) rectangle (4.5,1.5);
	\node[right=2pt] (t1) at (3.4,1.3) {rPST};
	\node[right=2pt] (t2) at (3.4,0.85) {rPLT};
	\node[right=2pt] (t3) at (3.4,0.4) {rHST};
	
	\fill[white] (5.5,0.2) rectangle (6.5,1.5);
	\node[right=2pt] (t4) at (5.4,1.3) {rHLT};
	\node[right=2pt] (t5) at (5.4,0.85) {bWST};
	\node[right=2pt] (t6) at (5.4,0.4) {bWLT};
	
	\fill[white] (7.55,0.2) rectangle (8.55,1.5);
	\node[right=2pt] (t4) at (7.5,1.3) {$0.85$};
	\node[right=2pt] (t5) at (7.5,0.85) {$0.97$};
	\node[right=2pt] (t6) at (7.5,0.4) {$1.0$};
\end{tikzpicture}
\caption{Forward analysis of the taste vector models on filtered listening history data.}
\label{fig:forward}
\end{figure}

In the forward analysis we take the first 100 songs of a user's listening history, which we use to generate the taste vector.
This taste vector is then compared to the next 50 song vectors in the listening history in terms of cosine distance.
That is, for each user $u$ in the test set we calculate the sequence $\left( L_{\cos}(\bvec{t}^u, \bvec{s}^u_{100+j}) \right)$, for $j \in \set{1, 2, \dots, 50}$.
Figure \ref{fig:forward} shows a plot of these sequences averaged over all users in the test set.
The overall trend of every model is that the cosine distance increases if we compare the taste vector to songs further in the future.
This is not surprising since it is generally easier to predict nearby songs than it is to predict songs in the far future, because the former are usually more related to the last played songs.
We see that the $\gamma=0.85$ and rPST model have comparable performance. They have low cosine distance for the first few tracks, but this quickly starts to rise, and they both become the worst performing models for long-term prediction.
All other models, apart from rHST and rHLT, behave similarly, with rPLT being slightly better and the $\gamma=0.97$ model slightly worse than all others.
The two best performing models are rHST and rHLT.
Until future track 20, the rHST model gives the lowest cosine distance, and rHLT is significantly the best model after that.
Since playlists are typically more coherent than listening histories---e.g.~they often contain entire albums or sometimes only songs by the same artist---this can explain why the playlist-trained RNNs, and especially rPST, perform not that well in this analysis.
Another general trend is that ST models typically perform better than their LT counterparts in the very near future.
And at some point the LT model becomes better than the ST model and is a better predictor on the long term.
Finally, among all baselines, we also observe that bWST is the best performing short-term model, and bWLT performs best to predict on the long term, which is not surprising since the weight-based models are a generalization of the discounting models.
Note that the classification models remain absent, because in this case the output of the RNN is not a user taste vector.

\subsubsection*{Backwards analysis}
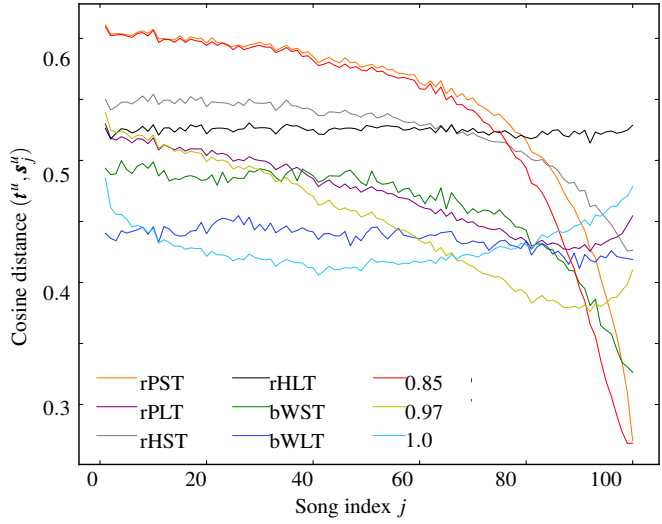
\begin{figure}[t!]
\begin{tikzpicture}[scale=.85,inner sep=0pt]
	\node [draw=none, anchor=south west] () at (0,0) {\includegraphics[trim = 3.1cm 2.5cm 2.5cm 2.0cm, clip, width=.465\linewidth]{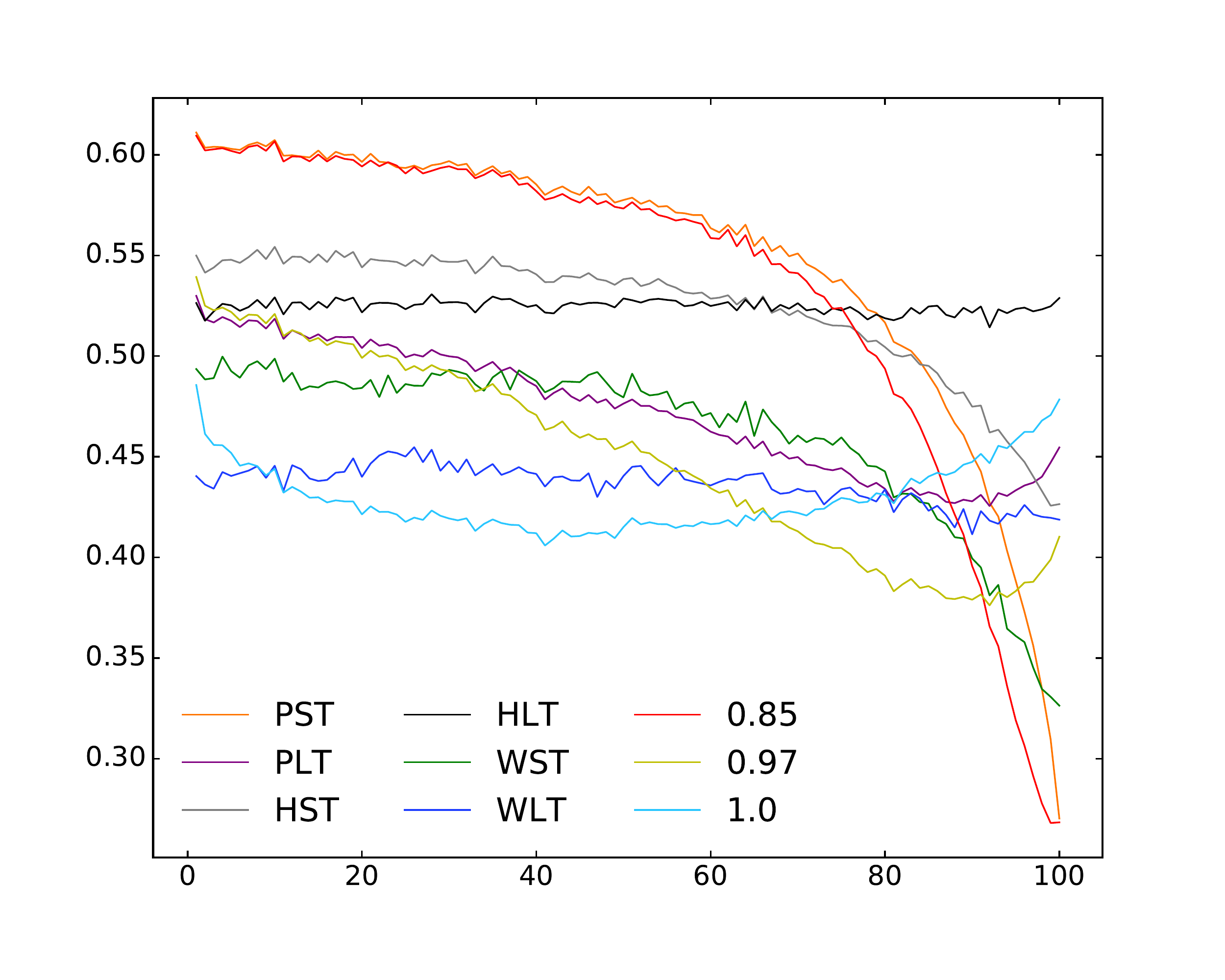}};

	\node[below=2pt] (bO) at (0.25,0.0) {$0$};
	\node[below=2pt] (b1) at (1.85,0.0) {$20$};
	\node[below=2pt] (b2) at (3.5,0.0) {${40}$};
	\node[below=2pt] (b3) at (5.05,0.0) {${60}$};
	\node[below=2pt] (b4) at (6.65,0.0) {${80}$};
	\node[below=2pt] (b5) at (8.20,0.0) {${100}$};
	\draw[draw opacity=0] (0.1,0.0) -- node[below=12pt] {Song index $j$} (8.4,0.0);
	\draw[draw opacity=0] (0.0, 0.1) -- node[left=15pt] {\rotatebox{90}{Cosine distance $(\bvec{t}^u, \bvec{s}^u_{j})$}} (0.0, 6.75);
	\node[left=2pt] (lO) at (0.0,0.9) {$0.3$};
	\node[left=2pt] (l1) at (0.0,2.7) {$0.4$};
	\node[left=2pt] (l2) at (0.0,4.55) {$0.5$};
	\node[left=2pt] (l3) at (0.0,6.35) {$0.6$};
	
	\fill[white] (1.0,0.2) rectangle (2.0,1.5);
	\node[right=2pt] (t1) at (0.9,1.3) {rPST};
	\node[right=2pt] (t2) at (0.9,0.85) {rPLT};
	\node[right=2pt] (t3) at (0.9,0.4) {rHST};
	
	\fill[white] (3.0,0.2) rectangle (4.0,1.5);
	\node[right=2pt] (t4) at (2.9,1.3) {rHLT};
	\node[right=2pt] (t5) at (2.9,0.85) {bWST};
	\node[right=2pt] (t6) at (2.9,0.4) {bWLT};
	
	\fill[white] (5.1,0.2) rectangle (6.1,1.5);
	\node[right=2pt] (t4) at (5.0,1.3) {$0.85$};
	\node[right=2pt] (t5) at (5.0,0.85) {$0.97$};
	\node[right=2pt] (t6) at (5.0,0.4) {$1.0$};
\end{tikzpicture}
\caption{Backwards analysis of the taste vector models on filtered listening history data.}
\label{fig:backward}
\end{figure}

In this analysis we again take the first 100 songs of a user' listening history, which we use to generate a taste vector.
We then compare this taste vector to these first 100 songs, i.e.~the songs that generated the taste vector.
We thus look back in time to gain insights as to what parts of the listening history contribute most or least to the taste vector.
For this we calculate the sequence $\left( L_{\cos}(\bvec{t}^u, \bvec{s}^u_{j}) \right)$, for $j \in \set{1, 2, \dots, 100}$, and Figure \ref{fig:backward} plots this sequence for each model averaged over all users in the test set.
We see that the rPST and $\gamma=0.85$ models are very focused on the last songs that were played, and the average cosine distance increases rapidly the further we go back in history: for songs 1 until 80 they are the worst performing.
These models will typically be very near-sighted in their predictions, that is, the song recommendations will mostly be based on the last 10 played tracks.
This is again due to the fact that playlists are very coherent, and predicting a near-future track can be done by looking at the last tracks alone.
The rHST and bWST models also show a similar behavior, but the difference in cosine distance for tracks in the near and far history is not as large compared to rPST.
The listening history RNNs, both rHST and rHLT, produce an overall high cosine distance.
These models are therefore not really tied to or focused on particular songs in the user's history.
It is interesting to note that the plot for rHLT and bWLT is a near-flat line, so that the produced taste vector lies equally far from all songs in terms of cosine distance.
In comparison, the $\gamma=1.0$ taste vector, which is actually just a sum of all songs, produces a U-shaped plot, which is a behavior similar to the long-term weights in Figure \ref{fig:weight_plot}.
If we would attribute more weight to the first and last few tracks, we would end up with a flatter line.
The $\gamma=0.97$ plot also has a U-shape, but the minimum is shifted more towards the recent listening history.
Note again that the classification models are absent in this analysis for the same reason as specified above.

\subsubsection*{Precision@k}
In this section we calculate the precision of actual song recommendations.
We again take 100 songs from a user's history which we use to generate a taste vector.
Then, as described in Section \ref{subsec:recommending}, we query the word2vec space for the $k$ nearest songs in the catalog in terms of cosine distance.
We will denote the resulting set as $\Omega\left( \bvec{t}^u, k \right)$.
The $precision@k$ value is then the fraction of how many songs in $\Omega\left( \bvec{t}^u, k \right)$ actually appear in the user's next $k$ tracks:
\begin{align}
&precision@k =\\
&\qquad\frac{1}{|U|}\sum_{u\in U} \frac{\left| \set{s\colon s\in \Omega\left( \bvec{t}^u, k \right) \wedge s\in (s^u_{101:100+k})} \right|}{k}.\nonumber
\end{align}
We can also generalize this to $precision@[j:k]$:
\begin{align}
&precision@[j:k] =\\
&\quad\frac{1}{|U|}\sum_{u\in U} \frac{\left| \set{s\colon s\in \Omega\left( \bvec{t}^u, k-j+1 \right) \wedge s\in (s^u_{100+j:100+k})} \right|}{k-j+1}.\nonumber
\end{align}
Here we disregard the user's first next $j-1$ tracks, since it is often easier to predict the immediate next tracks than it is to predict tracks further in the future.
For the next results, we also slightly alter the definition of $\Omega\left( \bvec{t}^u, k \right)$: given the fact that no song in $(s^u_{101:150})$ occurs in $(s^u_{1:100})$ for all users $u$, as described in Section \ref{sec:datafiltering}, we only regard the $k$ nearest songs of $\bvec{t}^u$ that do not appear in $(s^u_{1:100})$.
For the classification models, we simply take the top $k$ songs with the highest scores, and compare them to the ground truth.
If we denote these top $k$ songs by $\Omega\left( \mathcal{R}\left( \bvec{s}^u_{1:k}; \bmat{W} \right), k \right)$, we can reuse the same definition of $precision@k$ from above.

The results of the experiments are shown in Table \ref{table:precision}.
In bold we mark the best performing model for each task, and the second best model is underlined.
The overall precisions are quite low, but given that we aggressively filtered the listening data (Section \ref{sec:datafiltering}), the task is rather difficult.
The history-based RNNs clearly perform best in all tasks.
Generally, for $precision@10$, $@25$ and $@50$ all short-term models outperform the long-term models.
But once we skip the first 25 songs, which are easier to predict, the long-term models take over, which shows that listening behavior indeed changes over time.
The performance of the playlist RNNs and weight-based models are comparable to the exponential discount models, which we already saw in Figure \ref{fig:forward}.

All four classification models that we trained achieved the same precision score of 0 percent, so we listed them as one entry in Table \ref{table:precision}.
They were not able to correctly guess any of the 50 future songs a user might listen to.
We can think of many reasons why we see this result.
First, the output dimensionality of 6 million is extremely large, which makes it difficult to discriminate between different but comparable items.
The RecSys Challenge 2015 dataset, used in both the works of Hidasi et al.~and Tan et al., only has around 37,500 items to predict, which is 160 times less than 6 million.
Second, the number of weights in the classification model is orders of magnitudes larger than for the regression model, which causes learning to be much harder.
And third, the data in the works by Hidasi et al.~and Tan et al.~comes from user sessions, which are mostly contained and coherent sequences of songs a user listens to within a certain time span, see also Section \ref{sec:motivation}.
The listening history dataset in our work goes across user sessions to be able to recommend on the long term, which makes it much more difficult to model the temporal dynamics.
This is reflected in the overall low precision accuracies.

\subsubsection*{A note on scalability}
As indicated in the title of this article, our methodology should allow for scalable music discovery.
The training procedure---i.e.~training word2vec and the RNN---is not time-critical and can be trained offline.
Despite the fact that these procedures could be parallelized when needed \cite{Ji:2016va}, we will focus on the recommendation part of the system itself, which is more time-critical.

Since every user can be treated independently, the entire pipeline we have proposed in Figure \ref{fig:overview} is 'embarrassingly parallel' and can therefore be scaled up to as many computational nodes as available.
Retrieving song vectors comes down to a dictionary lookup in constant time $\BigO{1}$.
Calculating user taste vectors through the RNN is linear $\BigO{n}$ in the number of historical song vectors $n$ we consider.
An extensive study of the scalability of Annoy, the last part in the pipeline, is beyond the scope of this paper, and poses a trade-off between accuracy and performance: more tree nodes inspected leads generally to more accurate nearest neighbors, but a slower retrieval time (approximately linear in the number of inspected nodes)\footnote{There is an excellent web article by Radim Rehurek from 2014 which studies this in depth, see \url{rare-technologies.com/performance-shootout-of-nearest-neighbours-querying}}.
Retrieving $1{,}000$ nearest neighbors using 10 trees and $10{,}000$ inspected nodes only takes on average 2.6ms on our system, which is in same order of magnitude compared to the RNN prediction times given in Table \ref{table:compare-rnn-types}.

Combining all the above, sampling a taste vector from the rHLT RNN and retrieving the top 50 closest songs from the Annoy LSH tree over 1,000 runs takes on average 58ms on our system, while retrieving the top 50 songs from the BPR RNN takes on average 754ms, which is 13 times slower.

\begin{table}[t!]
\small
\caption{Results for the $\mathbf{precision@k}$ experiments on filtered listening history data.}
\label{table:precision}
\tabcolsep=0.23cm
\begin{tabular}{l | c c c c c}
\toprule
				& \multicolumn{5}{c}{Precision (\%)} \\
				& $@10$ & $@25$ & $@50$ & $@[25:50]$ & $@[30:50]$\\
\hline
rPST				& 1.64 			& 2.39 			& 2.81 			& 1.15 			& 0.94				\\
rPLT				& 1.27 			& 1.98 			& 2.64 			& 1.30 			& 1.06				\\
rHST				& \textbf{2.03}	& \textbf{2.95}	& \textbf{3.72}	& \uline{1.85} 	& \uline{1.53}		\\
rHLT				& 1.40 			& 2.31 			& \uline{3.25}	& \textbf{1.89} 	& \textbf{1.63}		\\
bWST				& 1.67 			& 2.37 			& 2.94			& 1.28 			& 1.04				\\
bWLT				& 1.32 			& 1.95 			& 2.62			& 1.31 			& 1.06				\\
$\gamma=0.85$	& \uline{1.94} 	& \uline{2.63} 	& 3.00			& 1.20 			& 0.96				\\
$\gamma=0.97$	& 1.56 			& 2.20 			& 2.77			& 1.30 			& 1.05				\\
$\gamma=1.0$		& 1.16 			& 1.78 			& 2.41			& 1.24 			& 1.02				\\
Classification  & 0.00           & 0.00          & 0.00          	& 0.00          	& 0.00\\
\bottomrule
\end{tabular}
\end{table}

\section{Conclusions}
\label{sec:conclusions}
We modeled users on large-scale online music streaming platforms for the purpose of new music discovery.
We sequentially processed a user's listening history using recurrent neural networks in order to predict a song he or she will listen to in the future.
For this we treated the problem as a regression rather than classification task, in which we predict continuous-valued vectors instead of distinct classes.
We designed a short-term and long-term prediction model, and we trained both versions on playlist data as well as filtered user listening history data.
The best performing models were chosen to be as small and efficient as possible in order to fit in large-scale production environments.
Incorporating extra play context features did not significantly improve the models.
We performed a set of experimental analyses for which we conclude that the history-based models outperform the playlist-based and all baseline models, and we especially pointed out the advantages of using the regression approach over the classification baseline models.
We also saw that there is indeed a difference between short-term and long-term listening behavior.
In this work we modeled these with different models.
One possible line of future work would be to design a single sequence-to-sequence model that captures both short and long term time dependencies to predict the entire future listening sequence \cite{Sutskever:2014ty}.

\section*{Acknowledgments}
Cedric De Boom is funded by a PhD grant of the Research Foundation - Flanders (FWO).
We greatly thank Nvidia for its donation of a Tesla K40 and Titan X GPU to support the research of the IDLab group at Ghent University.

\bibliographystyle{spmpsci}      

\begin{thebibliography}{10}
\providecommand{\url}[1]{{#1}}
\providecommand{\urlprefix}{URL }
\expandafter\ifx\csname urlstyle\endcsname\relax
  \providecommand{\doi}[1]{DOI~\discretionary{}{}{}#1}\else
  \providecommand{\doi}{DOI~\discretionary{}{}{}\begingroup
  \urlstyle{rm}\Url}\fi

\bibitem{AlRfou:2016uc}
Al-Rfou, R., Alain, G., Almahairi, A., al, e.: {Theano - A Python framework for
  fast computation of mathematical expressions.}
\newblock arXiv.org  (2016)

\bibitem{Bahdanau:2015vz}
Bahdanau, D., Cho, K., Bengio, Y.: {Neural Machine Translation by Jointly
  Learning to Align and Translate}.
\newblock In: ICLR (2015)

\bibitem{Barkan:2016wm}
Barkan, O., Koenigstein, N.: {Item2vec - Neural Item Embedding for
  Collaborative Filtering.}
\newblock RecSys Posters  (2016)

\bibitem{Bennett:2007uv}
Bennett, J., Lanning, S.: {The Netflix Prize}.
\newblock In: Proceedings of KDD cup and workshop (2007)

\bibitem{Charikar:2002km}
Charikar, M.: {Similarity estimation techniques from rounding algorithms.}
\newblock In: STOC (2002)

\bibitem{Chung:2014wf}
Chung, J., G{\"u}l{\c c}ehre, {\c C}., Cho, K., Bengio, Y.: {Empirical
  Evaluation of Gated Recurrent Neural Networks on Sequence Modeling.}
\newblock arXiv.org  (2014)

\bibitem{Collobert:2011tk}
Collobert, R., Weston, J., Bottou, L., Karlen, M., Kavukcuoglu, K., Kuksa, P.:
  {Natural Language Processing (Almost) from Scratch}.
\newblock The Journal of Machine Learning Research  (2011)

\bibitem{DeBoom:2016gm}
De~Boom, C., Van~Canneyt, S., Demeester, T., Dhoedt, B.: {Representation
  learning for very short texts using weighted word embedding aggregation}.
\newblock Pattern Recognition Letters  (2016)

\bibitem{Dror:2011bb}
Dror, G., Koenigstein, N., Koren, Y.: {Yahoo! music recommendations - modeling
  music ratings with temporal dynamics and item taxonomy.}
\newblock In: RecSys (2011)

\bibitem{Figueiredo:2016ut}
Figueiredo, F., Ribeiro, B., Faloutsos, C., Andrade, N., Almeida, J.M.: {Mining
  Online Music Listening Trajectories.}
\newblock In: ISMIR (2016)

\bibitem{Goodman:2001dp}
Goodman, J.: {Classes for fast maximum entropy training}.
\newblock In: 2001 IEEE International Conference on Acoustics, Speech, and
  Signal Processing. Proceedings (Cat. No.01CH37221 (2001)

\bibitem{Graves:2014vz}
Graves, A., Jaitly, N.: {Towards End-To-End Speech Recognition with Recurrent
  Neural Networks.}
\newblock In: ICML (2014)

\bibitem{Greff:2015wv}
Greff, K., Srivastava, R.K., Koutn{\'\i}k, J., Steunebrink, B.R., Schmidhuber,
  J.: {LSTM: A Search Space Odyssey}.
\newblock arXiv.org  (2015)

\bibitem{Hidasi:2015uq}
Hidasi, B., Karatzoglou, A., Baltrunas, L., Tikk, D.: {Session-based
  Recommendations with Recurrent Neural Networks.}
\newblock arXiv.org  (2016)

\bibitem{Hill:2016wg}
Hill, F., Cho, K., Korhonen, A., Bengio, Y.: {Learning to Understand Phrases by
  Embedding the Dictionary.}
\newblock TACL  (2016)

\bibitem{Hochreiter:1997fq}
Hochreiter, S., Schmidhuber, J.: {Long short-term memory}.
\newblock Neural Computation  (1997)

\bibitem{Hu:2008el}
Hu, Y., Koren, Y., Volinsky, C.: {Collaborative filtering for implicit feedback
  datasets}.
\newblock 2008 Eighth IEEE International {\ldots}  (2008)

\bibitem{Ji:2016va}
Ji, S., Satish, N., Li, S., Dubey, P.: {Parallelizing Word2Vec in Multi-Core
  and Many-Core Architectures.}
\newblock CoRR  (2016)

\bibitem{Johnson:2014tf}
Johnson, C.C.: {Logistic matrix factorization for implicit feedback data}.
\newblock In: NIPS 2014 Workshop on Distributed Machine Learning {\ldots}
  (2014)

\bibitem{Karpathy:2015wu}
Karpathy, A., Johnson, J., Fei-Fei, L.: {Visualizing and Understanding
  Recurrent Networks}.
\newblock arXiv.org  (2015)

\bibitem{Kingma:2015ku}
Kingma, D., Ba, J.: {Adam: A Method for Stochastic Optimization}.
\newblock In: ICLR (2015)

\bibitem{Koren:2009jg}
Koren, Y., Bell, R., Volinsky, C.: {Matrix Factorization Techniques for
  Recommender Systems}.
\newblock Computer  (2009)

\bibitem{Lee:2000ti}
Lee, D.D., Seung, H.S.: {Algorithms for Non-negative Matrix Factorization.}
\newblock NIPS 2000  (2000)

\bibitem{Liang:2016wj}
Liang, D., Altosaar, J., Charlin, L.: {Factorization Meets the Item Embedding:
  Regularizing Matrix Factorization with Item Co-occurrence}.
\newblock In: ICML Workshop (2016)

\bibitem{Maas:2013tn}
Maas, A.L., Hannun, A.Y., Ng, A.Y.: {Rectifier nonlinearities improve neural
  network acoustic models}.
\newblock In: ICML (2013)

\bibitem{Mikolov:2013uz}
Mikolov, T., Sutskever, I., Chen, K., Corrado, G., Dean, J.: {Distributed
  Representations of Words and Phrases and their Compositionality}.
\newblock In: NIPS 2013: Advances in neural information processing systems
  (2013)

\bibitem{Moore:2013tj}
Moore, J.L., Chen, S., Turnbull, D., Joachims, T.: {Taste Over Time - The
  Temporal Dynamics of User Preferences.}
\newblock In: ISMIR (2013)

\bibitem{Ozsoy:2016tm}
Ozsoy, M.G.: {From Word Embeddings to Item Recommendation}
\newblock arXiv.org  (2016)

\bibitem{Pan:2008kb}
Pan, R., Zhou, Y., Cao, B., Liu, N.N., Lukose, R., Scholz, M., Yang, Q.:
  {One-Class Collaborative Filtering}.
\newblock In: 2008 Eighth IEEE International Conference on Data Mining (2008)

\bibitem{Anonymous:H3H5BbuI}
Paterek, A.: {Improving regularized singular value decomposition for collaborative
  filtering}.
\newblock In: KDDCup 2007 (2007)

\bibitem{Rendle:2009wp}
Rendle, S., Freudenthaler, C., Gantner, Z., Schmidt-Thieme, L.: {BPR - Bayesian
  Personalized Ranking from Implicit Feedback.}
\newblock UAI  (2009)

\bibitem{dosSantos:2014tr}
dos Santos, C.N., Gatti, M.: {Deep Convolutional Neural Networks for Sentiment
  Analysis of Short Texts}.
\newblock In: COLING 2014, the 25th International Conference on Computational
  Linguistics (2014)

\bibitem{Sercu:2016ub}
Sercu, T., Goel, V.: {Advances in Very Deep Convolutional Neural Networks for
  LVCSR.}
\newblock In: Interspeech (2016)

\bibitem{Sutskever:2014ty}
Sutskever, I., Vinyals, O., Le, Q.V.: {Sequence to Sequence Learning with
  Neural Networks}.
\newblock In: NIPS 2014 (2014)

\bibitem{Tan:2016vy}
Tan, Y.K., Xu, X., Liu, Y.: {Improved Recurrent Neural Networks for
  Session-based Recommendations.}
\newblock arXiv.org  (2016)

\bibitem{VanDenOord:2013tp}
Van Den~Oord, A., Dieleman, S., Schrauwen, B.: {Deep content-based music
  recommendation}.
\newblock In: NIPS (2013)

\bibitem{VanDenOord:2016uo}
Van Den~Oord, A., Dieleman, S., Zen, H., Simonyan, K., Vinyals, O., Graves, A.,
  Kalchbrenner, N., Senior, A., Kavukcuoglu, K.: {WaveNet: A Generative Model
  for Raw Audio}.
\newblock arXiv.org  (2016)

\end{thebibliography}

\newpage
\appendix

\section{Table of Symbols}
In order of appearance:\\
\label{sec:appendix}
\begin{tabularx}{\textwidth}{l | X}
\toprule
$\vec{v}_u(t)$				& User vector for user $u$ at time $t$\\
$\vec{v}_i(t)$				& Item vector for item $i$ at time $t$\\
$r_{ui}$						& rating of item $i$ by user $u$\\
$\mu$						& Global average rating\\
$b_u(t)$						& Rating bias of user $u$ at time $t$\\
$b_i(t)$						& Rating bias of item $i$ at time $t$\\
$h_t	$						& Hidden state at time $t$\\
$c_t$						& Cell state at time $t$\\
$f_t$						& Forget gate at time $t$\\
$o_t	$						& Output gate at time $t$\\
$r_t$						& Reset gate at time $t$\\
$u_t$						& Update gate at time $t$\\
$U_x, W_x$					& Weight matrices for gate $x$\\
$w_x$						& Weight vector for gate $x$\\
$b_x$						& Bias for gate $x$\\
$\mathcal{F}(\cdot)$			& Non-linear function\\
$\sigma(\cdot)$				& Sigmoid function\\
$\odot$						& Element-wise multiplication operator\\
$N$							& Number of songs in the catalog\\
$D$							& Embedding dimensionality\\
$U$							& Set of all users on the platform\\
$\left(\vec{s}^u\right)$		& Ordered sequence of song vectors user $u$ listened to\\
$\vec{t}^u$					& Taste vector of user $u$\\
$\mathcal{R}\left( \cdot; \bmat{W} \right)$	& RNN function with parameters $\bmat{W}$\\
$\mathcal{L}(\cdot)$			& Loss function\\
$\norm{\cdot}_2$				& L2 norm\\
$L_{\cos}(\cdot)$			& Cosine distance\\
$\mathrm{unif}\set{x,y}$		& Uniform distribution between $x$ and $y$\\
$\mathcal{D}$				& Dataset of song sequences\\
$\ell_{\min}, \ell_{\max}$	& Minimum and maximum sampling offsets\\
$\eta$						& Learning rate\\
$\vec{c}^u$					& Context vector for user $u$\\
$\oplus$						& Vector concatenation operator\\
$C$							& Ordered set of contexts on the Spotify platform\\
$C_i$						& $i$'th context in $C$\\
$c(s)$						& set of contexts for song $s$\\
$\mathrm{onehot}(i, L)$		& One-hot vector of length $L$ with a 1 at position $i$\\
$\mathbf{1}_{A}(x)$			& Indicator function: 1 if $x\in A$, else 0\\
$\Delta(x, y)$				& Time difference between playing songs $x$ and $y$\\
$D_{hid}$					& Hidden dimensionality\\
$\gamma$						& Discount factor\\
$\mathcal{W}(\cdot; \vec{w})$	& Weight-based model function with weights $\vec{w}$\\
$\lambda$					& Regularization term\\
$\zeta(\cdot)$				& Riemann zeta function\\
$\mathrm{Zipf}_z(\cdot)$		& Zipf probability density function with parameter $z$\\
$rPST,\; rPLT$				& Short- and long-term playlist RNN\\
$rHST,\; rHLT$				& Short- and long-term user listening history RNN\\
$bWST,\; bWLT$				& Short- and long-term weight-based model\\
\bottomrule
\end{tabularx}

\end{document}